# ENHANCED SECURE KEY EXCHANGE SYSTEMS BASED ON THE JOHNSON-NOISE SCHEME

## Laszlo Bela Kish

*Texas A&M University, Department of Electrical and Computer Engineering, College Station, TX 77843-3128, USA;*
*email: Laszlo.Kish@ece.tamu.edu*

**Abstract**

We introduce seven new versions of the Kirchhoff-Law-Johnson-(like)-Noise (KLJN) classical physical secure key exchange scheme and a new transient protocol for practically-perfect security. While these practical improvements offer progressively enhanced security and/or speed for the non-ideal conditions, the fundamental physical laws providing the security remain the same.

In the "intelligent" KLJN (iKLJN) scheme, Alice and Bob utilize the fact that they exactly know not only their own resistor value but also the stochastic time function of their own noise, which they generate before feeding it into the loop. By using this extra information, they can reduce the duration of exchanging a single bit and in this way they achieve not only higher speed but also an enhanced security because Eve's information will significantly be reduced due to smaller statistics.

In the "multiple" KLJN (MKLJN) system, Alice and Bob have publicly known identical sets of different resistors with a proper, publicly known truth table about the bit-interpretation of their combination. In this new situation, for Eve to succeed, it is not enough to find out which end has the higher resistor. Eve must exactly identify the actual resistor values at both sides.

In the "keyed" KLJN (KKLJN) system, by using secure communication with a formerly shared key, Alice and Bob share a proper time-dependent truth table for the bit-interpretation of the resistor situation for each secure bit exchange step during generating the next key. In this new situation, for Eve to succeed, it is not enough to find out the resistor values at the two ends. Eve must also know the former key.

The remaining four KLJN schemes are the combinations of the above protocols to synergically enhance the security properties. These are: the "intelligent-multiple" (iMKLJN), the "intelligent-keyed" (iKKLJN), the "keyed-multiple" (KMKLJN) and the "intelligent-keyed-multiple" (iKMKLJN) KLJN key exchange systems.

Finally, we introduce a new transient-protocol offering practically-perfect security without privacy amplification, which is not needed at practical applications but it is shown for the sake of ongoing discussions.

## 1. Introduction

In this section we briefly define our basic terms of secure key exchange utilizing the laws of physics and introduce the Kirchhoff-Law-Johnson-(like)-Noise (KLJN) secure key exchange protocol.

*1.1 Conditional, Unconditional, Perfectly and Imperfectly secure key exchange*

In private-key based secure communication, the two parties Alice (A) and Bob (B) possess an identical secure key, which is not known to the public, and they are utilizing this key in a cipher software to encrypt/decrypt the messages they send/receive [1]. Thus, to able to communicate, Alice and Bob must first generate and share a secure key, which is typically a random bit sequence. The first important problem of secure communication is how to generate and share this key in a *secure* way between Alice and Bob. Note, even if Alice and Bob may already have shared a former secret key to communicate securely and they maybe able to share a new key via that secure communication, that is not a *security-growing* method because, if the old key is cracked by an eavesdropper (Eve), the new key will also become compromised implying that such a simple method cannot be used to share the new key. Of course, Alice and Bob may exchange a secure key by personally meeting or using a mail courier service however that is neither satisfactory for high speed nor secure enough (against spying) if they share many keys for future use. It is the safest to generate the new key "ad-hoc" when it is needed. Today's internet-based secure communications use software tools to generate and share secure keys where the reason why Eve (who is monitoring the channel) cannot extract the key is her limited computational power. However, *the whole information about the secure key is publicly accessible in the during the key exchange* [1]. Thus, these methods offer only conditional security because, with sufficient computing power (for example by having a hypothetical quantum computer or its noise-based-logic version), the key would instantaneously be cracked by Eve. Due to the unexpected progression of computing technologies, this type of security is not only conditional but also it is *not a future-proof-security* [2]: Eve can potentially crack the recorded key exchange and communication in the near future even if presently such task looks hopeless.

    Due to these facts, scientists have been exploring various physical phenomena for secure key exchange where the laws of physics could offer the security. The goal is to have a key exchange where either the exchange cannot be measured/recorded, or when the information measured/recorded by Eve is zero; a situation called perfect information theoretic security; or this information is practically miniscule, a situation is *practically-perfect* information theoretic security. If the extracted information by Eve is zero or small and, *its amount does not depend on Eve's accessible resources when she is approaching the limits imposed by the laws of physics*, the security is called *unconditional* [2,3]. In practical cases "unconditional" and "information theoretic" security are interchangeable terms [1]. Thus the security classification can be classified as *perfect unconditional* (perfect information theoretic), *imperfect unconditional* (imperfect information theoretic) or conditional, where "perfect" with practical elements can only be "practically perfect", and only simplified mathematical models of system may offer really perfect security at the theoretical level. With other words, perfect unconditional security can only be defined at the (simplified) conceptual level in any physical system while imperfect unconditional security is the level that *any* real physical system can aim to reach due to the limitations posed by non-ideal elements and situations [2,3]. Perfect security is like infinity, it can be approached but never reached.

*1.2 The Kirchhoff-Law-Johnson-(like)-Noise (KLJN) secure key exchange scheme*

The KLJN scheme is a statistical/physical competitor to quantum communicators and its security is based on Kirchhoff's Loop Law and the Fluctuation-Dissipation Theorem. More generally, it is founded on the Second Law of Thermodynamics, which indicates that the security of the ideal scheme is as strong as the impossibility to build a perpetual motion machine of the second kind.

Until 2005, it was a commonly accepted that only quantum key distribution (QKD) is able to perform information theoretic (unconditional) secure key exchange and that can theoretically provide perfect security while practically it is always imperfect. However, in 2005, the Kirchhoff-Law-Johnson-(like)-Noise (KLJN) secure key exchange [2-16] scheme was introduced [4] and later it was built and its security demonstrated [7]. These ideas have inspired new concepts also in computing, particularly noise-based logic and computing [17-24], where not the security of data but complexity of data processing has been the issue.

The core KLJN system, without the defense circuitry (current-voltage measurement/comparison, filters, etc.) against invasive and non-ideality attacks, is shown in Fig. 1.

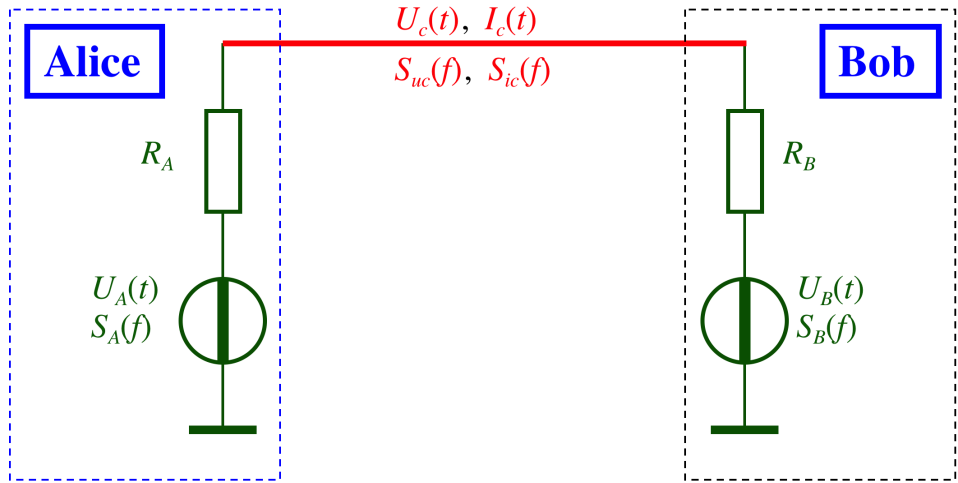

Fig. 1. Outline of the core KLJN system. Parasitic elements leading to non-ideal features and defense circuit block (current/voltage monitoring/comparison) against invasive attack are not the topics of this paper thus they are not shown/discussed here. The resistors $R_A$ and $R_B$ are randomly selected from the $\{R_L, R_H\}$ set.

We first briefly survey the foundations of the ideal KLJN system [2,4,9]. Fig. 1 shows a model of the idealized KLJN scheme designed for secure key exchange [4]. At each KLJN-clock period, which is the duration of a single bit exchange, Alice and Bob connect their randomly chosen resistor, $R_A$ and $R_B$, respectively, to the line. These resistors are randomly selected from the set $\{R_L, R_H\}$, $R_L \neq R_H$, where the elements represent the low, *L* (0), and high, *H* (1), bits, respectively. Alice and Bob randomly choose one of the resistors and connect it to the wire line. The situations *LH* and *HL* represent secure bit exchange [4], because Eve cannot distinguish between them through measurements, while *LL* and *HH* are insecure. The Gaussian voltage noise generators—delivering white noise with publicly agreed bandwidth—represent an enhanced thermal (Johnson) noise at a publicly agreed effective temperature $T_{eff}$ (typically $T_{eff} \geq 10^9 K$ [7]) where their noises are statistically independent from each other or from the noise during the former KLJN-clock period. According to the Fluctuation-Dissipation Theorem, the power density spectra $S_{u,L}(f)$ and $S_{u,H}(f)$ of the voltages $U_{L,A}(t)$ and $U_{L,B}(t)$ supplied by the voltage generators in $R_L$ and $R_H$ are given by

$$S_{u,L}(f) = 4kT_{eff}R_L \text{ and } S_{u,H}(f) = 4kT_{eff}R_H, \tag{1}$$

respectively.

In the case of secure bit exchange (*i.e.*, the *LH* or *HL* situation), the power density spectrum $S(f)$ and the mean-square amplitude $\langle U_{ch}^2 \rangle$ of the channel voltage $U_{ch}(t)$ and the same measures of the channel current $I_{ch}(t)$ are given as

$$\langle U_{ch,HL/LH}^2 \rangle = \Delta f \, S_{u,ch,HL/LH}(f) = 4kT_{eff} \frac{R_L R_H}{R_L + R_H} \Delta f \, , \qquad (2)$$

and

$$\langle I_{ch,HL/LH}^2 \rangle = \Delta f \, S_{i,ch,HL/LH}(t) = \frac{4kT_{eff}}{R_L + R_H} \Delta f \, , \qquad (3)$$

respectively, where $\Delta f$ is the noise-bandwidth; and further details are given elsewhere [4,9]. It should be observed that during the *LH* and *HL* cases, due to linear superposition, the spectrum given by Eq. (2) represents the sum of the spectra at two particular situations, *i.e.*, when only the noise generator of $R_L$ is running one gets

$$S_{L,u,ch}(f) = 4kT_{eff} R_L \left( \frac{R_H}{R_L + R_H} \right)^2 \, , \qquad (4)$$

and when the noise generator of $R_H$ is running one has

$$S_{H,u,ch}(f) = 4kT_{eff} R_H \left( \frac{R_L}{R_L + R_H} \right)^2 \, . \qquad (5)$$

The ultimate security of the system against passive attacks is provided by the fact that the power $P_{H \to L}$, by which the noise generator of resistor $R_H$ is heating resistor $R_L$, is equal to the power $P_{L \to H}$ by which the noise generator of resistor $R_L$ is heating resistor $R_H$ [4,9]. Thus the net power flow between Alice and Bob is zero, which is required by the Second Law of Thermodynamics. A proof of this can easily be derived from Eqs. (4,5) for the noise-bandwidth of $\Delta f$ :

$$P_{L \to H} = \frac{S_{L,u,ch}(f) \Delta f}{R_H} = 4kT_{eff} \frac{R_L R_H}{(R_L + R_H)^2} \, , \qquad (6a)$$

and

$$P_{H \to L} = \frac{S_{H,u,ch}(f) \Delta f}{R_L} = 4kT_{eff} \frac{R_L R_H}{(R_L + R_H)^2} \, . \qquad (6b)$$

The equality $P_{H \to L} = P_{L \to H}$ (*cf.* Eq. (6)) is in accordance with the Second Law of Thermodynamics; violating this equality would mean not only going against basic laws of physics (the inability to build a perpetual motion machine) but also allowing Eve to use the voltage-current cross-correlation $\langle U_{ch}(t) \, I_{ch}(t) \rangle$ to extract the bit [4]. However the only quantity that could provide directional information is zero, $\langle U_{ch,HL/LH}(t) \, I_{ch,HL/LH}(t) \rangle = 0$, and hence Eve has no information to determine the bit location during the *LH* and *HL* situations. This security proof against passive (listening) attacks holds only for Gaussian noise, which has the well-known property that its power density spectrum or autocorrelation function already provides the maximum achievable information about the noise, and no higher order distribution functions or other tools (such as higher-order statistics) are able to serve with additional information.

The error probability of the bit exchange between Alice and Bob is determined by the following issue. In the case of the *LL* bit status of Alice and Bob, which is not secure situation, the channel voltage and current satisfy:

$$\langle U^2_{ch,LL} \rangle = \Delta f \, S_{u,ch,LL}(f) = 4kT_{eff} \frac{R_L}{2} \Delta f \quad \text{and} \quad \langle I^2_{ch,LL} \rangle = \Delta f \, S_{i,ch,LL}(t) = \frac{2kT_{eff}}{R_L} \Delta f \quad , \qquad (7)$$

while, in the case of the other non-secure situation, the *HH* bit status, the channel voltage and current satisfy:

$$\langle U^2_{ch,HH} \rangle = \Delta f \, S_{u,ch,HH}(f) = 4kT_{eff} \frac{R_H}{2} \Delta f \quad \text{and} \quad \langle I^2_{ch,HH} \rangle = \Delta f \, S_{i,ch,HH}(t) = \frac{2kT_{eff}}{R_H} \Delta f \qquad (8)$$

During key exchange in this classical way, Alice and Bob must compare the predictions of Eqs. (2,3,7,8) with the actually measured mean-square channel voltage and current to decide if the situation is secure (LH or HL) while utilizing the fact that these mean-square values are different in each of these three situations (LL, LH/HL and HH). If the situation is secure, Alice and Bob will know that the other party has the inverse of his/her bit, which means, a secure key exchange takes place. To make an error-free key exchange, Alice and Bob must use a sufficiently large statistics, which means long-enough KLJN-clock period. However, the length of the KLJN-clock period determines the speed of the exchange of the whole key and, most importantly, Eve's statistics when utilizing non-ideal features to extract information. Thus protocols that can reduce the necessary duration of the KLJN-clock period for a satisfactory statistics (fidelity) for Alice and Bob would enhance not only the speed but also the security. The new ("intelligent") KLJN protocol described in Sec. 2 offers this kind of improvement.

### *1.3 On invasive attacks and non-idealities*

It should be observed [2,3,4,6,9,10,11,12] that deviations from the shown circuitry—including invasive attacks by Eve, parasitic elements, delay effects, inaccuracies, non-Gaussianity of the noise, *etc.*—will cause a potential information leak toward Eve. The circuit symbol "line" in the circuitry represents an ideal wire with uniform instantaneous voltage and current along it. Fortunately the KLJN system is very simple, implying that the number of such attacks is strongly limited. The defense method against attacks utilizing these aspects is straightforward and it is generally based on the comparison of instantaneous voltage and current data at the two ends via an authenticated communication between Alice and Bob.

These attacks are not subject of the present paper and we refer to our relevant papers where they have been analyzed [2,3,4,6,9,10,11,12] and misconceptions (or errors) of other papers written by others corrected. In surveys [2,3], existing invasive attacks by other authors and us have been surveyed.

It is important to emphasize that, if the security of a certain bit is compromised, that is *known also by Alice and Bob* therefore they can decide to discard the bit to have a clean secure key. The price for that is a reduced speed of key exchange however the perfect` security can be maintained. Alternatively privacy amplification can be executed [13], which also results in a slower key exchange. In conclusion, in the KLJN system, Alice and Bob can always protect themselves against eavesdropping of the key but they are still vulnerable against jamming the key exchange by Eve (the same situation exists in QKD, too, because the single photons are the most easy objects to jam).

## 2. The "intelligent" KLJN (iKLJN) key exchange protocol

The important characteristics of all passive attack types against practical KLJN systems is that Eve's bit-guessing success rate is strongly limited by poor statistics [4,6,9,10,11,12]. In most cases, Eve has a very weak signal-to-noise ratio due to the limited KLJN-clock period, which is the time window to make that statistics [11,12]. Thus, if we could further limit Eve's time window her success rate would further decrease. As we have already pointed out above, the minimum duration of the clock period in the original KLJN scheme is set by Alice and Bob by their need to successfully classify the measured mean-square channel voltage and/or current levels by comparing them with the predictions of Eqs. (2,3,7,8) in order to identify that to which one of the LL, LH/HL, HH situations does the actual status corresponds [4].

The Intelligent KLJN (iKLJN) system allows using shorter KLJN-clock period thus it further weakens Eve's statistics. It has the same hardware as the original KLJN system but the protocol is more calculation-intensive. Alice and Bob utilize the fact that they exactly know not only their own resistor value but also the stochastic time function of their own noise, which they generate before feeding it into the loop. In the iKLJN method, Alice and Bob, by utilizing the superposition theorem on the channel noise, subtract their own contribution to generate a *reduced-channel-noise* that does not contain their own noise component. Because they don't know the resistance value at the other end, they must run two alternative *computational-schemes* simultaneously to calculate the reduced-channel-noises to account for the possible resistance situations (totally four time functions, two voltage and two current noises corresponding to the two possible resistance situations at the other end), see below. Then they analyze that at which one of these situations the reduced-channel-noise does not contain their noise contribution. The reduced-channel-noise that does not contain their noise component has been calculated with the correct assumption about the actual resistance value used by the other party. Thus the nature of the decision Alice and Bob makes is changed in the iKLJN: instead of evaluating mean-square noise amplitudes, they must assess the independence of two noise processes. Note, obviously they continue to assess the channel noise situation also in the classical way by evaluating the mean-square of the cannel noise amplitudes, which has partially independent information, thus combining the new and old information sources in the guessing process significantly shortens the KLJN-clock period needed for a given error probability.

At the same time, Eve can only use her old way, the parasitic elements to extract any information. Because Eve's available observation time window (the KLJN-clock period set by Alice and Bob) becomes shorter, the information that she can extract is also significantly reduced. She may not even be sure during the shortened KLJN-clock period if a secure bit exchange took place, or not, therefore her related error rate will increase. Thus, in the non-ideal situation, when information-leak exists, the reduced observation time window progressively worsens Eve's probability of successfully guessing not only the key bits but also guessing which KLJN-clock periods had secure bit exchange.

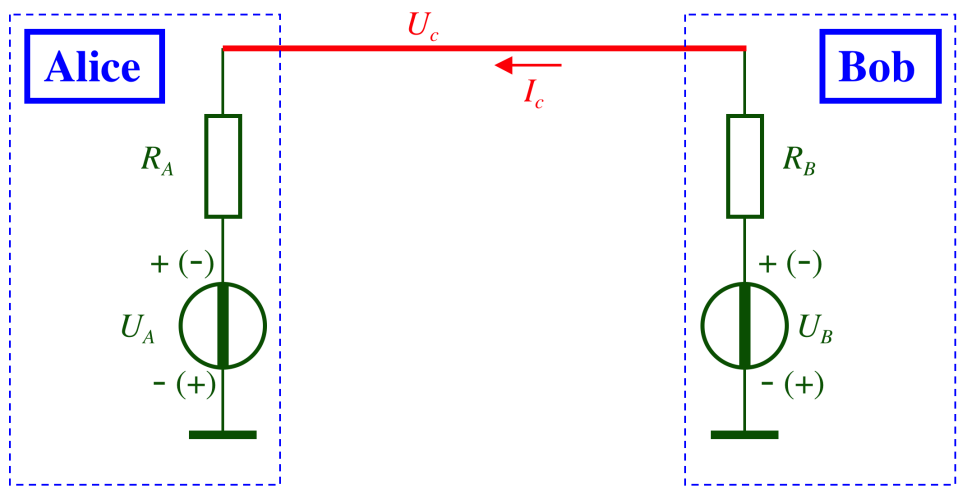

Fig. 2. Snapshot of the current and the voltages in the KLJN system: at a given time moment, the polarities of Alice's and Bob's voltages and the resulting current are shown.

## 2.1 Analysis of the "intelligent" KLJN (iKLJN) key exchange protocol

To analyze the system with this new approach and to illustrate its way of functioning, first, let us assume that Bob's resistance is $R_B$ and Alice's one is:

$$R_A = \alpha R_B \tag{9}$$

where $\alpha \neq 1$. We analyze Bob's protocol to demonstrate the process. Alice is acting in a similar way, which results in the same type of features.

According to Kirchhoff's Loop Law, the channel noise current $I_c(t)$ and noise voltage $U_c(t)$ at a given instant of time are given as:

$$I_c = \frac{U_B - U_A}{R_B(1+\alpha)} \tag{10}$$

$$U_c = \frac{U_A + \alpha U_B}{1+\alpha}, \tag{11}$$

where, for convenience, we skipped the time variable from the equations. Bob's calculation of the reduced-channel-noise currents and voltages is carried out in the following way.

*Hypothesis-1: Reduced-channel-noise with the incorrect assumption*

In one of the computational-schemes, Bob supposes that the resistance value of Alice is the same as his one, that is, $R_A = R_B$, which is the incorrect assumption. Then the "incorrect" reduced-channel-current and reduced-channel-voltage amplitudes, $I_{c,1}^*$ and $U_{c,1}^*$, are:

$$I_{c,1}^* = I_c - \frac{U_B}{2R_B} = \frac{U_B - U_A}{R_B(1+\alpha)} - \frac{U_B}{2R_B} = -\frac{1}{R_B}\frac{2U_A - U_B(1-\alpha)}{2(1+\alpha)} \tag{12}$$

$$U_{c,1}^* = U_c - \frac{U_B}{2} = \frac{U_A + \alpha U_B}{1+\alpha} - \frac{U_B}{2} = \frac{2U_A - U_B(1-\alpha)}{2(1+\alpha)} \tag{13}$$

*Hypothesis-2: Reduced-channel-noise with the correct assumption*

In the other computational-scheme, Bob supposes that the resistance value of Alice is different than his one, that is, $R_A = \alpha R_B$, which is the correct assumption. Then the "correct" reduced-channel-current and reduced-channel-voltage amplitudes, $I_{c,2}^*$ and $U_{c,2}^*$, are:

$$I_{c,2}^* = I_c - \frac{U_B}{R_B(1+\alpha)} = \frac{U_B - U_A}{R_B(1+\alpha)} - \frac{U_B}{R_B(1+\alpha)} = \frac{-U_A}{R_B(1+\alpha)} \tag{14}$$

$$U_{c,2}^* = U_c - \frac{\alpha U_B}{1+\alpha} = \frac{U_A + \alpha U_B}{1+\alpha} - \frac{\alpha U_B}{\alpha+1} = \frac{U_A}{1+\alpha} \tag{15}$$

It is obvious from our approach and the results in Eqs (12-15) that, in the case of the incorrect assumption, the reduced-channel-noises contain both the noise contribution of Alice ($U_A(t)$) and that of Bob ($U_B(t)$) while, in the case of the incorrect assumption, they contain the noise of Alice ($U_A(t)$) only. Thus, Bob, by using a proper statistical tool to compare the reduced-channel-noises with his own noise ($U_B(t)$) and checking for the independence, he can identify the "correct" assumption and, in this way, learn the actual resistor value of Alice. One of the simplest ways of that is the evaluation of the cross-correlations between his noises and the reduced-channel-noises:

With the incorrect assumption:

$$\langle U_B I_{c,1}^* \rangle = \frac{U_B^2(1-\alpha)}{2(1+\alpha)R_B} \neq 0 \tag{16}$$

$$\langle U_B U_{c,1}^* \rangle = \frac{\langle U_B^2 \rangle(\alpha-1)}{2(1+\alpha)} \neq 0 \tag{17}$$

With the correct assumption:

$$\langle U_B I_{c,2}^* \rangle = \frac{-\langle U_A U_B \rangle}{R_B(1+\alpha)} = 0 \qquad (18)$$

$$\langle U_B U_{c,2}^* \rangle = \frac{\langle U_A U_B \rangle}{1+\alpha} = 0 \qquad (19)$$

While evaluating and comparing these cross-correlations is a new and independent source of information for Bob about Alice's resistance; in the field of statistics, there are more efficient ways of estimations, which will be explored in the future.

Naturally, Alice is proceeding in the same way as Bob. Then combining this "intelligent" method with the original assessments based on the voltage and current noises will significantly enhance the information of Alice and Bob, and thus they can use shorter KLJN-clock period to reach the same error probability as earlier. And, because Eve does not have access to the "intelligent" way of information extraction, her successful guessing probability of the resistor situation in the non-ideal KLJN system will drop significantly compared to the original KLJn situation, while in the ideal system her information remains zero. Moreover, due to the reduction of KLJN-clock duration, also Eve's error probability of guessing if a secure bit exchange took place or not will decrease, which further enhances her uncertainty.

Finally, an important question: *What is the price of this "intelligent" enhancement of the original KLJN protocol*? A higher computational capacity is needed for Alice and Bob to carry out this task, which implies higher electrical power requirements, too. Thus, when computational performance is limited or low power requirements are essential, the classical KLJN protocol (and its "multiple" and/or "keyed" versions, see below) is the way to go.

## 3. An educational problem about the KLJN system

This section is not essential for the rest of the paper thus Readers who are not interested in the deeper understanding of the foundations of the KLJN system can jump to the next section. A natural question arises about the iKLJN theory: *Do we need statistical evaluation or perhaps we could quickly determine the correct assumption by just using the reduced voltage and current values*? Similar questions arose in 2005 before Johnson (-like) noise was introduced into the KLJN system and a random DC voltage generator pair version of it was explored. The answer was that Johnson noise and statistical analysis were needed. That study was unpublished and, we will now show that the similar situation in the iKLJN system leads to the same conclusion; statistics cannot be avoided.

If we use Ohm's law between the correctly deduced reduced-channel-voltage and reduced-channel-current components, we get

$$\frac{U_{c,1}^*}{I_{c,1}^*} = -\frac{1}{R_B} \quad , \qquad (20)$$

which is the expected result where the negative sign is due to the direction of current component into Bobs resistor from Alice's voltage generator, see Fig 2.

Our naive expectation may be that, if we do the same derivation with the *incorrectly deduced* current and voltage components then the result will be different and then Bob can instantaneously find out that *Hypothesis-2* is the valid assumption for Alice's resistor.

Unfortunately, even the incorrect assumption yields the same result:

$$\frac{U^*_{c,2}}{I^*_{c,2}} = -\frac{1}{R_B} \tag{21}$$

This surprising result is due to the degeneracy of the system of equations describing the channel voltage and current, which also prohibits for Bob to deduce Alice's resistor even though he knows his resistor and voltage (a situation that led the author in 2005 to test thermal noise in this system). No "instantaneous" way to find out Alice's resistance, that is the value of $\alpha$, exists due to this degeneracy. In conclusion, only statistical methods can provide the necessary information for Bob and Alice.

## 4. The "multiple" KLJN (MKLJN) key exchange protocol

In the "multiple" KLJN (MKLJN) system, Alice and Bob have publicly known identical sets of different resistors $\{R_1 < R_2 < ... < R_n\}$. For each KLJN-clock period, they randomly choose a resistor from this set and connect it (with a corresponding independent noise generator) to the line. There is a publicly known truth table about the bit-interpretation of the different combinations of the chosen $[R_i, R_j]$ resistor pair, whenever $R_i \neq R_j$, with the condition that the bit interpretation of $[R_i, R_j]$ is the inverse of the bit interpretation of $[R_j, R_i]$. It is designed so that, when the estimation of one of the resistors is missed at one of the sides and the neighboring resistor value is estimated instead, the bit-interpretation reverses in order to make Eve's guessing statistics worse.

In this new situation, for Eve to succeed, it is *not enough* to find out which end has the higher resistance. Eve must *exactly identify* the actual resistor values at both sides (while Alice and Bob only at the other side) to know that which $[R_i, R_j]/[R_j, R_i]$ situation is the relevant in the truth table and, in accordance with the original KLJN principle, even then Eve is unable to decide if $[R_i, R_j]$ or $[R_j, R_i]$ is the case. The result of modification is again an enhanced security in the non-ideal case.

## 5. The "keyed" KLJN (KKLJN) key exchange protocol

This enhancement is inspired by Horace Yuen's "keyed" quantum key exchange (called KCQ) [25] to enhance the security of his new quantum key exchange protocol. It works after a secure key is already generated/shared by Alice and Bob in the KLJN protocol. Then, by using secure communication with the shared key, they share a time-dependent truth table for the bit-interpretation of the $[R_L, R_H]_i$ versus $[R_H, R_L]_i$ resistor situation at the *i*-th secure bit exchange step during generating the next key. Note, the $[R_L, R_H]_i$ and $[R_H, R_L]_i$ situations must always mean opposite bit values.

It is obvious that KKLJN is a security growing technique because, even if Eve would succeed with correctly guessing the former key and learn the truth table, the security of the new key is still information theoretical (unconditional) and it only "falls back" to the security level of the original KLJN key exchange. In the non-ideal case, if Eve has no information about the former key, the information of Eve about the key is progressively less than at the standard KLJN and in the ideal case her information remains zero.

## 6. The "keyed-multiple" KLJN (KMKLJN) key exchange protocol

Naturally, the KKLJN protocol can be enriched by using multiple resistor sets, $n > 2$, in the same fashion as the MKLJN system is doing but, instead of a publicly known truth table the bit-interpretation of the $[R_L, R_H]_i$ versus $[R_H, R_L]_i$ resistor situations is randomly changed for the subsequent key exchanges and the relevant truth table is shared by secure communication utilizing the former key. The KMKLJN protocol synergically combines the security enhancement of the MKLJN and KKLJN protocols.

## 7. Three more protocols: iMKLJN, iKKLJN and iKMKLJN

The "multiple", "keyed" and "keyed-multiple" KLJN protocols can be combined with the "intelligent" method of accessing the resistors at the other end by Alice and Bob to reduce the KLJN-clock duration and Eve's information, and to increase the speed. This enrichment should always be made whenever calculation power is enough. These are: the "intelligent-multiple" (iMKLJN), the "intelligent-keyed" (iKKLJN), the "keyed-multiple" (KMKLJN) and the "intelligent-keyed-multiple" (iKMKLJN) KLJN key exchange systems.

## 8. Transient-protocol for practically-perfect security

Finally, we introduce a new transient-protocol offering *practically-perfect* security without privacy amplification, which is not needed at practical applications but shown for the sake of ongoing discussions. At the beginning of each bit-exchange period (KLJN-clock period), until the thermalization, which is the mixing and equilibration of noise-propagation (not waves) in the cable takes place, there is a potential for information leak. Even though, so far, no concrete method has been show that would be able to extract the key bit, precautions have been made to reduce this effect by low-pass line-filters (which always must be present to keep away wave-generating frequency components of the wave-limit, and to prevent hacking attacks), ramping-up the generator voltages or, instead of ramping starting the generators from zero voltage. However, even in these cases, one can expect some small information leak at the beginning because the system starts with noises that have not been mixed. This information leak is supposed to be very small with these stochastic signals (still needs a concrete attack to see how much it is) because of the relaxation time constant of the low-pass filters is much longer than the propagation time in the cable (this is required by the no-wave limit). Thus, we show the "ultimate transient protocol" below for the sake of discussions not for practical applications.

The protocol is as follows. First Alice and Bob randomly decide if they want to use $R_L$ or $R_H$ during the next KLJN-clock period for their $R_A$ and $R_B$, respectively. Then, to execute the key exchange, they use continuously variable resistors (such as potentiometers, etc.). If noise generators also used to enhance the noise-temperature than the band-limited white noise spectra of the noise-generators are also variable in a synchronized fashion so that the noise-temperature stays constant, at the publicly agreed value $T_{eff}$. At the beginning of the KLJN-clock period, both Alice and Bob start with:

$$R_A(0) = R_B(0) = \frac{R_L + R_H}{2} \qquad (22)$$

and they stay at this value until the noises equilibrate in the wire. Thus no informative transients can be observed after connecting the resistors to the line. Then Alice and Bob execute independent slow continuous-time random-walks with their resistor values (and in a in a synchronized fashion with the spectral parameter of their noise generators). The random walks are executed so slowly that, from a thermodynamic point of view, the system is changing in the adiabatic limit: there is virtually a thermal equilibrium in the line during the whole random-walk process.

There is a publicly pre-agreed time period $t_r$ to execute these independent random walks. If within this time period Alice and Bob reach their randomly preselected $R_A$ and $R_B$ value, they stop the random walk and stay at this value. Then, after the $t_r$ time period, they start the KLJN protocol, in the regular way or a proper advanced fashion described in this paper. In this way, the transient effects and the information leak they may cause, are virtually kept away.

If by the end of the $t_r$ time period, either the random-walk of Alice or Bob (or both) does not reach the randomly preselected resistance value, he/she (or both) submits a cancellation signal via an authenticated channel and the bit exchange process is terminated and a new KLJN-clock period starts in the way described above.

Note, it is also possible to introduce alternative protocols where there is no preliminary random decision by Alice and Bob and they use the random value they get by the random walk at the end of the $t_r$ time period. If the difference of obtained resistance values is large enough they can use it for secure key exchange. The fact if the difference is large enough will turn out only at the end f the KLJN-clock period and then they can decide (based on publicly known rules) if they keep the exchanged bit.

## 9. Summary

While the KLJN secure key exchange has perfect security at the idealized (mathematical) conditions, in the case of non-ideal devices (including non-zero range with finite speed), there is some information leak, similarly to quantum key exchange. To reach *practically-perfect* security, Alice and Bob can do privacy amplification [13] or discarding the bits [9] that provide information to Eve beyond a certain threshold. In this paper, we show additional ways to enhance the security in the non-ideal cases by using one of the seven new extended protocols introduced above.

In Table 1, we summarize the hardware and computational requirements of the various protocols.

|  | KLJN | iKLJN | MKLJN | KKLJN | KMKLJN | iMKLJN | iKKLJN | iKMKLJN |
|---|---|---|---|---|---|---|---|---|
| Number $n$ of resistors | 2 | 2 | $n > 2$ | 2 | $n > 2$ | $n > 2$ | 2 | $n > 2$ |
| Loops to compute | - | 2 | - | - | - | $n(n-1)$ | 2 | $n(n-1)$ |

Table 1. Comparison of the standard KLJN hardware requirements with that of the improved versions.

It should be noted that the foundations of the security remain the same: the Second Law of Thermodynamics, the particular properties of Gaussian stochastic processes, and Kirchhoff's Law.

Finally, we should emphasize that, due to the exchange and comparison of the current/voltage data at the two ends of the line, Alice and Bob exactly know Eve's information [9]. This is a new and unique situation in physical secure key exchange systems because QKD does not have this advantage. Though the exact implications of this fact must yet to be explored, it is obvious that it can offer formerly unexpected ways for enhancing security by properly discarding risky bits.

**Acknowledgements**

Discussions with Horace Yuen about his KCQ system and other general issues of security, and with Robert Mingesz and Zoltan Gingl about the KLJN system are appreciated.